\documentclass[12pt]{article}
\usepackage[cp1251]{inputenc}
\usepackage{amsmath}
\usepackage{amsfonts}
\usepackage{amssymb}
\def\pa{\partial}                       
\def\beq{\begin{eqnarray}}    
\def\eeq{\end{eqnarray}}      

\begin{document}
\date{}
\begin{center}
{\Large\textbf{On interactions of massless spin 3  and
scalar  fields}}

\vspace{18mm}

{\large
P.M. Lavrov$^{(a, b)} \footnote{E-mail: lavrov@tspu.edu.ru}$
}

\vspace{8mm}

\noindent  ${{}^{(a)}} ${\em
Center of Theoretical Physics, \\
Tomsk State Pedagogical University,\\
Kievskaya St.\ 60, 634061 Tomsk, Russia}

\noindent  ${{}^{(b)}} ${\em
National Research Tomsk State  University,\\
Lenin Av.\ 36, 634050 Tomsk, Russia}

\vspace{20mm}

\begin{abstract}
\noindent
Using new approach for the deformation procedure in the case of  reducible gauge theories
(Lavrov in Eur. Phys. J. C 82:429, 2022),  it is shown that in the model of  massless  spin 3
field and a real scalar field local cubic and  quartic
vertices invariant under original gauge transformations can be  explicitly constructed.

\end{abstract}

\end{center}

\vfill

\\

\newpage

\section{Introduction}
The Batalin-Vilkovisky (BV) formalism \cite{BV,BV1} being the most powerful covariant
quantization method plays the very important role in solving the deformation problem
which is considered as a way in construction of suitable interactions among
fields.
 The reason is related with embedding the problem
into solutions to the classical master-equation of the BV-formalism. Standard approach
\cite{BH,H} operates  infinite number of equations appeared in expansion of the solution
with respect to the deformation parameter.
Then the system of these equations is analyzed in each order with respect to
the deformation parameter using cohomological methods.
Quite recently,
the new method to find solutions of the deformation procedure within the BV-formalism
has been proposed \cite{BL-1,BL-2,L-2022}.\footnote{Analog of this approach
in canonical formalism for dynamical systems with constraints within
the Batalin-Fradkin-Vilkovisky formalism \cite{BFV1,BFV2,BFV3}
is known as well \cite{BL-3}.}
The method is based on the invariance
of the classical master-equation under anticanonical transformations acting
in minimal antisymplectic space of the BV-formalism. The anticanonical transformations
(for recent developments see \cite{BLT,BLT-21})
by itself present an effective tool in studying principal properties of general gauge theories.
So, the gauge fixing procedure and the gauge dependence of Green functions
can be described in terms of special anticanonical transformations observed firstly
in \cite{VLT}. Moreover, the anticanonical transformations admit to describe the
renormalization procedure of general gauge theories \cite{VLT} that, in its turn, allowed to
introduce new understanding the renormalization problem beyond the usual index arguments
\cite{GWen}. Here, we meet with a new face of anticanonical transformations
in the solution of the deformation problem, which make it possible to represent
solutions in an explicit and closed form.

In the present paper, we apply the new method \cite{BL-1,BL-2,L-2022} to construct
 suitable interactions between massless spin 3 and scalar fields
as the result of deformation of initial free action. The initial action is sum of the
Fronsdal action for massless spin 3 fields \cite{Fronsdal-1} and the action for
a free real scalar field. The choice of initial model allows to use the BV-formalism and
the method \cite{L-2022} directly because the model belongs to the class of
reducible gauge theories. In the case of massless integer spin $s>3$ fields, the
direct application
of the BV-formalism is impossible due to the traceless conditions on fields. Fortunately,
the situation is not so hopeless because there exist  unconstrained formulations for higher
spin fields \cite{BGK,BG,BIZ} which make possible application of the BV-formalism
and  new method  in high spin theory.
In the case under consideration we study cubic and quartic vertices containing one spin 3
field and two or three scalar fields using special
anticanonical transformations acting non-trivially in the sector of spin 3 fields only.
It is shown that the gauge algebra does not deform under proposed transformations. It has
the following consequence: a) the cubic vertices appearing as a local part  of
the deformed action disappears under requirement of the gauge invariance, b) the local quartic
vertices invariant under the gauge transformations do exist. In particular,
it means appearance for the first time in quantum field theory an explicit and
consistent local gauge model
of interacting fields with spin $s>2$.

The paper is organized as follows. In section 2, a short presentation of new
approach to the deformation procedure in the BV-formalism for reducible gauge
theories is given. In section 3, the Fronsdal action for spin 3 fields as
a gauge action of the first-stage reducible theory  is discussed.
In section 4, deformations of free model of massless spin 3 and scalar fields
responsible for local cubic and quartic vertices are studied.
In section 5, we summarize the results.

We  use the DeWitt's condensed notations
\cite{DeWitt}
and employ the symbols $\varepsilon(A)$ for the Grassmann parity and
${\rm gh}(A)$ for the ghost number respectively. The right and left
functional derivatives are marked by special symbols $"\leftarrow"$
and $"\overrightarrow{}"$ respectively. Arguments of any functional
are enclosed in square brackets $[\;]$, and arguments of any
function are enclosed in parentheses, $(\;)$. The partial right derivative of a function
$F(A)$ with respect to field $A^i$ is denoted as $F_{,i}(A)$.

\section{Deformations in BV-formalism}
In this section, we consider in short main statements of the new approach
\cite{BL-1,BL-2,L-2022} in solving the deformation problem within the BV-formalism.
 Starting point
of the BV-formalism is a gauge theory of the fields $A=\{A^i\}$ with Grassmann
parities $\varepsilon(A^i)=\varepsilon_i$ and ghost numbers $\;{\rm
gh}(A^i)=0$. The theory is described by an initial action $S_0[A]$.
It is assumed that  the action is invariant under the
gauge transformations\footnote{To simplify presentation of all
relations containing the right functional derivative of functional
$S[A]$ with respect
to field $A^i$ we will use the symbol $S_{,i}[A]$ .}
\beq
\label{ggen0}
S_{0,i}[A]R^i_{\alpha}(A)=0, \quad \delta A^i=R^i_{\alpha}(A)\xi^{\alpha},
\quad \alpha = 1,2,..., m
\eeq
where $R^i_{\alpha}(A)$
($\varepsilon(R^i_{\alpha}(A))=\varepsilon_i+\varepsilon_{\alpha},\;
{\rm gh}(R^i_{\alpha}(A))=0$) are gauge generators, and  gauge parameters
$\xi^{\alpha}$ ($\varepsilon(\xi^{\alpha})=\varepsilon_{\alpha}$)
are arbitrary functions of space-time coordinates.
It is assumed that the fields $A=\{A^i\}$ are linear independent with
respect to the index $i$ however, in general, the generators $R^i_{\alpha}(A)$ may
be linear dependent with respect to index $\alpha$.
Linear dependence of $R^i_{\alpha}(A)$ implies that the matrix
 $R^i_{\alpha}(A)$ has at the extremals $S_{0,j}[A]=0$ zero-eigenvalue
 eigenvectors $Z^{\alpha}_{\alpha_1}=Z^{\alpha}_{\alpha_1}(A)$, such
 that
\begin{eqnarray}
\label{GTR1}
 R^i_{\alpha}(A)Z^{\alpha}_{\alpha_1}(A) =S_{0,j}[A]K^{ji}_{\alpha_1}(A),\quad
 \alpha_1 = 1, ..., m_1,
\end{eqnarray}
 and the number $\varepsilon_{\alpha_1}=0,1$ can be found in such a way
 that $\varepsilon(Z^{\alpha}_{\alpha_1})=\varepsilon_{\alpha}+
 \varepsilon_{\alpha_1}$. Matrices $K^{ij}_{\alpha_1}=K^{ij}_{\alpha_1}(A)$
 can be chosen to possess the properties:
\begin{eqnarray}
\nonumber
 K^{ij}_{\alpha_1}=-(-1)^{\varepsilon_i\varepsilon_j}K^{ji}_{\alpha_1}
 ,\quad
\varepsilon(K^{j\;\!i}_{\alpha_1}) =
\varepsilon_i+\varepsilon_j+\varepsilon_{\alpha_1}.
\end{eqnarray}
Here, we restrict ourself to the case of first-stage reducibility when
the set of zero-eigenvalue eigenvectors $Z^{\alpha}_{\alpha_1}$ is linear independent
with respect to the index $\alpha_1$.

In general, the generators $R^i_{\alpha}(A)$ satisfy the
following relations
\beq
\label{ga} R^i_{\alpha ,
j}(A)R^j_{\beta}(A)-(-1)^{\varepsilon_{\alpha}\varepsilon_
{\beta}}R^i_{\beta ,j}(A)R^j_{\alpha}(A)=-R^i_{\gamma}(A)
F^{\gamma}_{\alpha\beta}(A) - S_{0,j}[A]M^{ji}_{\alpha\beta}(A),
\eeq
where
$F^{\gamma}_{\alpha\beta}(A)=F^{\gamma}_{\alpha\beta}$
($\varepsilon(F^{\gamma}_{\alpha\beta})=\varepsilon_{\alpha}+
\varepsilon_{\beta}+\varepsilon_{\gamma},\;{\rm
gh}(F^{\gamma}_{\alpha\beta})=0$) are the structure coefficients which may depend
on the fields $A^i$. They obey  the following
symmetry properties $F^{\gamma}_{\alpha\beta}=
-(-1)^{{\varepsilon_{\alpha}\varepsilon_{\beta}}}F^{\gamma}_{\beta\alpha}$,
and $M^{ij}_{\alpha\beta}(A)=M^{ij}_{\alpha\beta}$ satisfy the conditions
$M^{ij}_{\alpha\beta} = -(-1)^{\varepsilon_i\varepsilon_j}
M^{ji}_{\alpha\beta} =
-(-1)^{\varepsilon_{\alpha}\varepsilon_{\beta}}M^{ij}_{\beta\alpha}$.
If $M^{ij}_{\alpha\beta}=0$ then the gauge algebra is called closed.

According to main statements of papers \cite{BL-1,BL-2,L-2022}, the deformation procedure for
general gauge theories can be described in terms of the set of generating functions $h^i(A)$
of anticanonical transformations
depending on initial fields $A^i$ only and having the same transformation properties
and quantum numbers as for $A^i$. The deformed action, $\widetilde{S}[A]$,
is defined by the relation
\beq
\widetilde{S}[A]=S_0[A+h(A)]
\eeq
being invariant under the deformed gauge transformation
\beq
\widetilde{S}_{,i}[A]\widetilde{R}^i_{\alpha}(A)=0,\quad
\widetilde{\delta}A=\widetilde{R}^i_{\alpha}(A)\xi^{\alpha},
\eeq
where $\widetilde{R}^i_{\alpha}(A)$ are the deformed gauge generators,
\beq
\widetilde{R}^i_{\alpha}(A)=(M^{-1}(A))^i_{\;j}R^j_{\alpha}(A+h(A)).
\eeq
Here, the matrix $(M^{-1}(A))^i_{\;j}$
is inverse to
\beq
\label{M}
M^i_{\;j}(A)=\delta^i_{\;j}+h^i_{\;,j}(A) ,
\eeq
The deformed gauge generators on extremals of the deformed action are linear dependent,
\beq
\widetilde{R}^i_{\alpha}(A)\widetilde{Z}^{\alpha}_{\alpha_1}(A) =
\widetilde{S}_{,j}[A]\widetilde{K}^{ji}_{\alpha_1}(A),
\eeq
where the functions $\widetilde{Z}^{\alpha}_{\alpha_1}(A)$ and
$\widetilde{K}^{ji}_{\alpha_1}(A)$ are
\beq
\widetilde{Z}^{\alpha}_{\alpha_1}(A)=Z^{\alpha}_{\alpha_1}(A+h(A)),\quad
\widetilde{K}^{ji}_{\alpha_1}(A)=-
(M^{-1}(A))^j_{\;l}(M^{-1}(A))^i_{\;k}K^{kl}_{\alpha_1}(A+h(A))
(-1)^{\varepsilon_l\varepsilon_i}.
\eeq
The deformed gauge generators satisfy the relations
\beq
\widetilde{R}^i_{\alpha ,
j}(A)\widetilde{R}^j_{\beta}(A)-(-1)^{\varepsilon_{\alpha}\varepsilon_
{\beta}}\widetilde{R}^i_{\beta ,j}(A)\widetilde{R}^j_{\alpha}(A)=
-\widetilde{R}^i_{\gamma}(A)
\widetilde{F}^{\gamma}_{\alpha\beta}(A)-\widetilde{S}_{,j}[A]
\widetilde{M}^{ji}_{\alpha\beta}(A),
\label{dga},
\eeq
where $\widetilde{F}^{\gamma}_{\alpha\beta}(A)=F^{\gamma}_{\alpha\beta}(A+h(A))$
and $\widetilde{M}^{ji}_{\alpha\beta}(A)=-
(M^{-1}(A))^j_{\;l}(M^{-1}(A))^i_{\;k}M^{kl}_{\alpha\beta}(A+h(A))
(-1)^{\varepsilon_l\varepsilon_i}$.
Therefore, the deformed theory looks like as the gauge theory of the first-stage reducibility
similar to the initial one. It is remarkable fact that the deformation of an initial
gauge theory
is described in the explicit and closed form with the help of generating functions $h^i(A)$ of
special anticanonical transformations in the BV-formalism. Moreover, all arbitrariness in
the deformation procedure is controlled by the generating functions $h^i(A)$ only. Any choice
of these functions guarantees that the deformed action will be invariant under the
corresponding deformed gauge transformations.

Deformation of initial action looks
like as the shift of argument $A\rightarrow A+h(A)$. Such kind of transformations is trivial
in the case of local function $h(A)$. To obtain non-trivial deformations, the function $h(A)$
must be non-local. It means the non-locality of the deformed action. It may happen that
there exists a local sector of the deformed action. If this local part of the deformed
action will be invariant
under local part of the deformed gauge generators then the deformation procedure creates
a consistent local gauge-invariant theory. Namely, in paper \cite{BL-1} it was exactly
shown that
the Yang-Mills theory has been reproduced by special non-local deformation
of free vector field model.

\section{Fronsdal action for massless  spin 3 fields}
\noindent
The Fronsdal  action for massless  spin 3 fields
has the form
\beq
\nonumber
&&S_0[\varphi]=\int dx\Big[\varphi^{\mu\nu\rho}(x)\Box\varphi_{\mu\nu\rho}(x)-
3\eta_{\mu\nu}\eta^{\rho\sigma}\varphi^{\mu\nu\delta}(x)
\Box\varphi_{\rho\sigma\delta}(x)-
\frac{3}{2}\varphi_{\mu\nu\lambda}\eta^{\mu\nu}\pa^{\lambda}\pa_{\alpha}
\varphi^{\alpha\beta\gamma}\eta_{\beta\gamma}-\\
\label{b1}
&&\qquad\qquad\qquad
-3\varphi^{\mu\rho\sigma}(x)\pa_\mu\pa^\nu\varphi_{\nu\rho\sigma}(x)+
6\eta_{\mu\nu}\varphi^{\mu\nu\delta}(x)\pa^\rho\pa^\sigma
\varphi_{\rho\sigma\delta}(x)\Big],
\eeq
where $\varphi^{\mu\nu\lambda}$ is completely symmetric third rank tensor,
$\Box$ is the D'Alembertian, $\Box=\pa_{\mu}\pa^{\mu}$, and $\eta_{\mu\nu}$ is
the metric tensor of flat Minkowski space of the dimension $d$.
The action (\ref{b1})
is invariant under the gauge transformations\footnote{The symbol $(\cdots)$ means
the cycle permutation of indexes involved.}
\beq
\label{b3}
\delta\varphi^{\mu\nu\lambda}=\pa^{(\mu}\xi^{\nu\lambda)}=
R^{\mu\nu\lambda}_{\rho\sigma}\xi^{\rho\sigma}
\eeq
when gauge parameters $\xi^{\mu\nu}$
are subjected to the traceless condition,
\beq
\label{b4}
\eta_{\mu\nu}\xi^{\mu\nu}=0.
\eeq
The gauge generators, $R^{\mu\nu\lambda}_{\rho\sigma}=\pa^{(\mu}
\delta^{\nu}_{\rho}\delta^{\lambda)}_{\sigma},$ do  not dependent on fields.
They are symmetric in lower indices,
$R^{\mu\nu\lambda}_{\rho\sigma}=R^{\mu\nu\lambda}_{\sigma\varrho},$
and, therefore, satisfy the relations
\beq
\label{b7}
R^{\mu\nu\lambda}_{\rho\sigma}Z^{\rho\sigma}_{\ell}=0,
\quad Z^{\rho\sigma}_{\ell}=-Z^{\sigma\rho}_{\ell},\quad
\ell=1,2,..., \frac{d(d-1)}{2},
\eeq
with  antisymmetric matrices $Z^{\rho\sigma}_{\ell}$
which are linear independent
with respect to index $\ell$.
We assume that $Z^{\rho\sigma}_{\ell}$ are
Lorentz second-rank tensor with respect to upper indices. Therefore, in
the BV-formalism we have to consider the case $s=3$ as a gauge theory with
first-stage reducible algebra specified by additional restrictions,
$F^{\gamma}_{\alpha\beta}(A)=0,\; M^{ij}_{\alpha\beta}(A)=0,\;K^{ij}_{\alpha_1}(A)=0$.
The zero-eigenvalue eigenvectors, $Z^{\rho\sigma}_{\ell}$,
do not depend on fields. In particular, it means that they do not change
in the deformation process.

\section{Deformations of initial action}
We are going to study a  possibility in construction of interactions of massless
spin 3 fields with a real scalar field $\phi=\phi(x)$. We assume that the initial action
has the form
\beq
\label{c1}
S_0[\varphi,\phi]=S_0[\varphi]+S_0[\phi],
\eeq
where $S_0[\varphi]$ is defined in (\ref{b1}) and $S_0[\phi]$ is the action of a free real
scalar field,
\beq
\label{c1a}
S_0[\phi]=\int dx\frac{1}{2}\big[ \pa_{\mu}\phi\;\pa^{\mu}\phi-m^2\phi^2\big] .
\eeq
The action (\ref{c1}) is invariant under the following gauge transformations
\beq
\label{c2}
\delta\varphi^{\mu\nu\lambda}=\pa^{(\mu}\xi^{\nu\lambda)},\quad
\delta\phi=0, \quad
\eta_{\mu\nu}\xi^{\mu\nu}=0,
\eeq
and, therefore, belongs to the class of first-stage reducible theories.
Relations with notations used in the Sec.2 are
\beq
\nonumber
&&A^i=(\varphi^{\mu\nu\lambda},\phi),\quad
R^i_{\alpha}(A)= (R^{\mu\nu\lambda}_{\rho\sigma}, 0),\quad
Z^{\alpha}_{\alpha_1}(A)=(Z^{\rho\sigma}_{\ell}, 0), \quad
\xi^{\alpha}=(\xi^{\rho\sigma}, 0),\\
&&
F^{\gamma}_{\alpha\beta}(A)=0,\quad M^{ij}_{\alpha\beta}(A)=0,\quad K^{ij}_{\alpha_1}(A)=0.
\eeq

Now, we consider possible deformations of initial action using the procedure which is ruled out
by the generating functions $h^i(A)$. Here, we restrict ourself by the case of
anticanonical transformations acting effectively in the sector of fields
$\varphi^{\mu\nu\lambda}$ of the initial theory. It means the following structure of
generating functions
$h^i(A)=(h^{\mu\nu\lambda}(\varphi,\phi), 0)$. In construction of suitable generating
functions
$h^{\mu\nu\lambda}=(h^{\mu\nu\lambda}(\varphi,\phi)$, we have to take
into account the dimensions of quantities involved in the initial action $S_0[\varphi,\phi]$,
\beq
\label{c3}
{\rm dim}(\varphi^{\mu\nu\lambda})={\rm dim}(\phi)=\frac{d-2}{2},\quad
{\rm dim}(\xi^{\mu\nu})=\frac{d-4}{2},\quad {\rm dim}(\pa_{\mu})=1, \quad
{\rm dim}(\Box)=2.
\eeq
The generating function $h^{\mu\nu\lambda}$ should be symmetric and non-local  with
the dimension equals to $(d-2)/2$.
The non-locality will be achieved by using the operator
$1/\Box$.\footnote{It seems quite natural to control non-locality with the help
of operator $1/\Box$ because the kinematic part of free models
contains terms $A\Box A$.  Existence of other forms of the non-locality
leading to  a local part of the deformed action remains open.}
To construct cubic vertices $\sim \varphi\phi\phi$,
the $h^{\mu\nu\lambda}$ should be
at least quadratic in fields $\phi$. The tensor structure of $h^{\mu\nu\lambda}$
is obeyed by using
partial derivatives $\pa_{\mu}$  and the metric tensor
$\eta_{\mu\nu}$. The minimal number of derivatives equals to 3. Therefore,
the more general form of
$h^{\mu\nu\lambda}=h^{\mu\nu\lambda}(\phi)$ satisfying requirements listed above reads
\beq
\nonumber
&&h^{\mu\nu\lambda}=a_0\frac{1}{\Box}
\big(c_1\pa^{\mu}\pa^{\nu}\pa^{\lambda}\phi \;\phi+
c_2\pa^{(\mu}\pa^{\nu}\phi\;\pa^{\lambda)}\phi+c_3\eta^{(\mu\nu}\Box\pa^{\lambda)}\phi\;\phi
+\\
\label{c4}
&&\qquad\qquad\quad+c_4\Box\phi\;\eta^{(\mu\nu}\pa^{\lambda)}\phi+
c_5\eta^{(\mu\nu}\pa^{\lambda)}\pa_{\sigma}\phi\;\pa^{\sigma}\phi\big),
\eeq
where $a_0$ is the coupling constant with ${\rm dim}(a_0)=-d/2$ and $c_i,\;\; i=1.2,...,5$
are arbitrary dimensionless constants.
Local part of the deformed action has the form
\beq
\label{c5}
S_{loc}[\varphi,\phi]=S_0[\varphi,\phi]+S_{1\; loc}[\varphi,\phi]
\eeq
where $S_{1\; loc}=S_{1\; loc}[\varphi,\phi]$,
\beq
\nonumber
&&S_{1\; loc}=2a_0\int dx \varphi_{\mu\nu\lambda}\Big[c_1
\pa^{\mu}\pa^{\nu}\pa^{\lambda}\phi \;\phi+
c_2\pa^{(\mu}\pa^{\nu}\phi\;\pa^{\lambda)}\phi-
\big(c_1+c_3(d+1)\big)\eta^{(\mu\nu}\Box\pa^{\lambda)}\phi\;\phi
-\\
\label{c6}
&&\qquad\qquad\qquad-\big(c_2+c_4(d+1)\big)\Box\phi\;\eta^{(\mu\nu}\pa^{\lambda)}\phi-
\big(2c_2+c_5(d+1)\big)\eta^{(\mu\nu}\pa^{\lambda)}
\pa_{\sigma}\phi\;\pa^{\sigma}\phi\Big],
\eeq
corresponds to possible cubic vertices. Due to the special structure of generating functions
$h^i(A)=(h^{\mu\nu\lambda}(\phi),0)$, the matrix $(M^{-1}(A))^i_{\;j}$
can be found in the explicit form
\beq
(M^{-1}(A))^i_{\;j}=\left(\begin{array}{cc}
                     E^{\mu\nu\lambda}_{\rho\sigma\gamma}&
                     -h^{\mu\nu\lambda}(\phi)\overleftarrow{\pa}_{\!\phi}\\
                     0& 1\\
                     \end{array}\right),
\eeq
where $E^{\mu\nu\lambda}_{\rho\sigma\gamma}$ are elements of the unit matrix in the space of
third rank symmetric tensors. Therefore, the deformed gauge generators coincide
with initial ones,
$\widetilde{R}^i_{\alpha}(A)=R^i_{\alpha}(A)$.

Consider the variation of $S_{1\; loc}$
under the gauge transformations (\ref{c2}). We have
\beq
\nonumber
&&\delta S_{1\; loc}=-6a_0\int dx \xi_{\nu\lambda}\Big[-
(c_1+2c_3(d+1))\Box \pa^{\nu}\pa^{\lambda}\phi\;\phi-\\
\nonumber
&&\qquad\qquad\qquad\quad-
(c_1+c_3(d+1)+2c_4(d+1))\Box\pa^{(\nu}\phi\;\pa^{\lambda)}\phi+\\
\nonumber
&&\qquad\qquad\qquad\quad+
(c_1-3c_2-2c_5(d+1))\pa_{\sigma}\pa^{\nu}\pa^{\lambda}\phi \;\pa^{\sigma}\phi
-(c_2+c_5(d+1))\pa_{\sigma}\pa^{(\nu}\phi\;\pa^{\lambda)}\pa^{\sigma}\phi-\\
\label{c13}
&&
\qquad\qquad\qquad\quad-(c_2+2c_4(d+1))\Box \phi\;\pa^{\nu}\pa^{\lambda}\phi\Big],
\eeq
where the constraint $\eta^{\mu\nu}\xi_{\mu\mu}=0$ was used.
The system of algebraic equations appears
\beq
\nonumber
&&c_1+2c_3(d+1)=0,\quad c_1-3c_2-2c_5(d+1)=0,\quad c_2+c_5(d+1)=0,\\
\label{c14}
&& c_1+c_3(d+1)+c_4(d+1)=0,\quad c_2+2c_4(d+1)=0
\eeq
as consequence of requirement $\delta S_{1 loc}=0$.
This system has the solution
\beq
\label{c15}
c_2=c_1,\quad c_3=c_4=\frac{1}{2}c_5=-\frac{1}{2(+1)}c_1,
\eeq
which leads to the important conclusion that the action
\beq
\nonumber
&&S_{1\; loc}=2a_0c_1\int dx \varphi_{\mu\nu\lambda}\Big[
\pa^{\mu}\pa^{\nu}\pa^{\lambda}\phi \;\phi+
\pa^{(\mu}\pa^{\nu}\phi\;\pa^{\lambda)}\phi+
\frac{1}{2}\eta^{(\mu\nu}\Box\pa^{\lambda)}\phi\;\phi
+\\
\label{c16}
&&\qquad\qquad\qquad\qquad\qquad\quad
+\frac{1}{2}\Box\phi\;\eta^{(\mu\nu}\pa^{\lambda)}\phi-
\eta^{(\mu\nu}\pa^{\lambda)}
\pa_{\sigma}\phi\;\pa^{\sigma}\phi\Big],
\eeq
describes cubic vertices invariant under original gauge transformations.

In connection with the obtained result, a natural question arises:
Do gauge invariant quartic vertices $\varphi\phi\phi\phi$ in the theory under consideration
exist?
Repeating the main arguments that led to the construction of the generating function
(\ref{c4}) of the anticanonical transformation, the most general form of
the generating function $h^{\mu\nu\lambda}$
with three derivatives responsible for the generation of quartic  vertices reads
\beq
\nonumber
&&h^{\mu\nu\lambda}=a_1\frac{1}{\Box}
\big[c_1\pa^{\mu}\pa^{\nu}\pa^{\lambda}\phi \;\phi^2+
c_2\pa^{(\mu}\pa^{\nu}\phi\;\pa^{\lambda)}\phi\;\phi+
c_3\pa^{\mu}\phi\;\pa^{\nu}\phi\;\pa^{\lambda}\phi+
c_4\eta^{(\mu\nu}\pa^{\lambda)}\Box\phi\;\phi^2
+\\
\label{c17}
&&\qquad\qquad\quad+
c_5\eta^{(\mu\nu}\pa^{\lambda)}\pa_{\sigma}\phi\;\pa^{\sigma}\phi\;\phi+
c_6\eta^{(\mu\nu}\pa^{\lambda)}\phi\;\pa_{\sigma}\phi\;\pa^{\sigma}\phi+
c_7\Box\phi\;\eta^{(\mu\nu}\pa^{\lambda)}\phi\;\phi\big],
\eeq
where $a_1$ is a coupling constant with ${\rm dim}(a_1)=-(d-1)$ and $c_i,\;\; i=1,2,...,7$
are arbitrary dimensionless constants. For the local addition,
$S_{2\;loc}=S_{2\;loc}[\varphi,\phi]$, to the initial action (\ref{c1}),
we obtain
\beq
\nonumber
&&S_{2\;loc}=2a_1\int dx\varphi_{\mu\nu\lambda}
\Big[c_1\pa^{\mu}\pa^{\nu}\pa^{\lambda}\phi \;\phi^2+
c_2\pa^{(\mu}\pa^{\nu}\phi\;\pa^{\lambda)}\phi\;\phi+
c_3\pa^{\mu}\phi\;\pa^{\nu}\phi\;\pa^{\lambda}\phi-\\
\nonumber
&&\qquad\qquad\quad-(c_1+c_4(d+1))\eta^{(\mu\nu}\pa^{\lambda)}\Box\phi\;\phi^2
-(2c_2+c_5(d+1))\eta^{(\mu\nu}\pa^{\lambda)}\pa_{\sigma}\phi\;\pa^{\sigma}\phi\;\phi-\\
\label{c18}
&&\qquad\qquad\quad-(c_3+c_6(d+1))\eta^{(\mu\nu}\pa^{\lambda)}\phi\;
\pa_{\sigma}\phi\;\pa^{\sigma}\phi-
(c_2+c_7(d+1))\Box\phi\;\eta^{(\mu\nu}\pa^{\lambda)}\phi\;\phi\Big].
\eeq
Notice that as in previous case, the gauge generators do not transform under anticanonical
transformations generated by functions (\ref{c17}).

Omitting the algebraic  manipulations, we find the variation $S_{2\;loc}$ under gauge
transformations (\ref{c2}) in the form
\beq
\nonumber
&&\delta S_{2\;loc}=6a_1\!\!\int \!\!dx \xi_{\nu\lambda}
\Big[(c_1+2c_4(d+1))\pa^{\nu}\pa^{\lambda}\Box \phi\;\phi^2\!-\!
(2c_1-3c_2-2c_5(d+1))\pa^{\sigma}\pa^{\nu}\pa^{\lambda}\phi\;\pa_{\sigma}\phi\;\phi\!+\!\\
\nonumber
&&\qquad\qquad+2(2c_1+2c_4(d+1)+c_7(d+1))\Box\pa^{\nu}\phi\;\pa^{\lambda}\phi\;\phi+
2(c_2+c_5(d+1))\pa^{\sigma}\pa^{\nu}\phi\;\pa^{\lambda}\pa_{\sigma}\phi\;\phi+\\
\nonumber
&&\qquad\qquad+(c_2+2c_7(d+1))\pa^{\nu}\pa^{\lambda}\phi\;\Box\phi\;\phi+
2(c_2+c_3+(c_5+2c_6)(d+1))\pa^{\sigma}\pa^{\nu}\phi\;\pa^{\lambda}\phi\;\pa_{\sigma}\phi-\\
\label{c21}
&&\qquad\qquad\!-\!(c_2\!-\!2c_3\!-\!2c_6(d+1))\pa^{\nu}
\pa^{\lambda}\phi\;\pa_{\sigma}\phi\pa^{\sigma}\!\phi\!-\!
(c_3\!-\!2c_2\!-\!2c_7(d\!+\!1))\Box\phi\;\pa^{\nu}\phi\;\pa^{\lambda}\phi
\Big].
\eeq
Note that the system  of algebraic equations
\beq
\nonumber
&&c_1+2c_4(d+1)=0,\quad 2c_1-3c_2-2c_5(d+1)=0,\quad
2c_1+2c_4(d+1)+c_7(d+1)=0,\\
\nonumber
&& c_2+c_5(d+1)=0,\quad
c_2+2c_7(d+1)=0,\quad c_2+c_3+(c_5+2c_6)(d+1)=0,\\
\label{c22}
&&
c_2-2c_3-2c_6(d+1)=0,\quad c_3-2c_2-2c_7(d+1)=0
\eeq
has non-trivial solution
\beq
\label{c22}
c_2=c_3=2c_1,\quad c_4=-\frac{1}{2(d+1)}c_1,\quad
c_5=-\frac{2}{d+1}c_1, \quad c_6=c_7=-\frac{1}{d+1}c_1 .
\eeq
Therefore, the functional
\beq
\nonumber
&&S_{2\;loc}=2a_1\!\int \!dx\varphi_{\mu\nu\lambda}
\Big[\pa^{\mu}\pa^{\nu}\pa^{\lambda}\phi \;\phi^2+
2\pa^{(\mu}\pa^{\nu}\phi\;\pa^{\lambda)}\phi\;\phi+
2\pa^{\mu}\phi\;\pa^{\nu}\phi\;\pa^{\lambda}\phi-
\frac{1}{2}\eta^{(\mu\nu}\pa^{\lambda)}\Box\phi\;\phi^2-\\
\label{c23}
&&\qquad\qquad\qquad
-2\eta^{(\mu\nu}\pa^{\lambda)}\pa_{\sigma}\phi\;\pa^{\sigma}\phi\;\phi-
\eta^{(\mu\nu}\pa^{\lambda)}\phi\;
\pa_{\sigma}\phi\;\pa^{\sigma}\phi-
\Box\phi\;\eta^{(\mu\nu}\pa^{\lambda)}\phi\;\phi\Big]
\eeq
is gauge invariant,
\beq
\delta S_{2\;loc}=0,
\eeq
and presents the quartic vertices.

The local action
\beq
\label{newmod}
S[\varphi,\phi]=S_0[\varphi,\phi]+S_{2\;loc}[\varphi,\phi]
\eeq
 describes the model of interacting
$\varphi^{\mu\nu\lambda}$ and $\phi$ fields which is invariant under gauge transformations
(\ref{c2}) and belongs to the class of first-stage reducible theories.
In fact, the model (\ref{c23}), (\ref{newmod})
is the first new example of local gauge theory beyond cubic interactions
 constructed with using the new method
\cite{BL-1,BL-2,L-2022}. Moreover, in quantum field
theory the action (\ref{newmod}) presents the first consistent local gauge model
of interacting fields with integer spin $s>2$.

\section{Discussion}
In the present paper, the new approach \cite{BL-1,BL-2,L-2022} to the deformation procedure
 \cite{BH,H} has been applied to construct local cubic and quartic vertices for massless spin
 3 and
scalar fields. It was shown that non-local anticanonical transformations
acting non-trivial only in the sector of spin 3 fields allow to exist  local
cubic vertices
being invariant under original gauge transformations while local gauge-invariant
quartic vertices do exist. In fact, for the first time, the  new gauge model of
interacting fields including fields with integer spin $s>2$ was explicitly constructed.
Let us note that reproducing the obtained result in the standard Noether procedure
adopted  for construction of interactions in the theory of higher spin fields \cite{BH,H}
is not possible without its essential reformulation.
Consider in more details relations between the standard
Noether procedure and the new method.
From the standard Noether procedure, the new method
looks formally like as an explicit summation of infinite Taylor series
in powers of the deformation parameter $g$.
Indeed, the deformed action $S=S[A]$  and deformed gauge transformation
$\delta_{\xi} A=R(A)\xi$ are presented as
\beq
S=S_0+gS_1+g^2S_2+\cdots ,\quad
\delta_{\xi} A=\delta_{\xi}^{(0)}A+g\delta_{\xi}^{(1)}A+
g^2\delta_{\xi}^{(2)}A+\cdots,
\label{D1}
\eeq
where $g$ is the deformation parameter,  $S_0=S_0[A]$ and $\delta_{\xi}^{(0)}A=R^{(0)}(A)\xi$
are the initial gauge action
and original gauge transformations, respectively.
The standard Noether procedure
to the deformation means the infinite set of coupled relations
appearing as consequence of the invariance of the deformed action under
deformed gauge transformations, $\delta_{\xi} S=0$,
\beq
\nonumber
&&\delta_{\xi}^{(1)}S_0+\delta_{\xi}^{(0)} S_1=0,\\
\label{D2}
&&\delta_{\xi}^{(2)}S_0+\delta_{\xi}^{(1)}S_1+\delta_{\xi}^{(0) }S_2=0,
\eeq
and so on.
In the new approach, firstly,
an explicit form both  the deformed action and deformed
gauge symmetry are found and, secondly, arbitrariness
in solutions is established. It is proved that the deformation
is described by special anticanonical transformations in the minimal
antisymplectic space of the BV formalism. The anticanonical transformations
by itself are defined by generating function $h(A)$ in a  number equal
to original fields $A$. In terms of $h(A)$ the deformed action $S=S[A]$ reads
\beq
\label{D3}
S[A]=S_0[A+gh(A)]
\eeq
 while the deformed gauge transformations have the following
closed form
\beq
\label{D4}
\delta_{\xi}A=M^{-1}(A)R^{(0)}(A+gh(A))\xi,
\eeq
where $M^{-1}(A)$ is
the matrix inverse to $M(A)=I+gh(A)\overleftarrow{\pa}_{\!A}$. The evident advantage of new
method follows from the two above statements. All process of deformation is controlled
by one function $h(A)$ of anticanonical transformation which is unique
arbitrariness appearing in
the new method.
For
any anticanonical transformation the deformed action obeys
invariance under corresponding deformed gauge transformation,
$\delta_{\xi}S[A]=0$. Now, solutions to the standard Noether procedure read
\beq
\nonumber
&&S_1[A]=S_0[A]\overleftarrow{\pa}_{\!A}h(A),\quad
S_2[A]=\frac{1}{2}S_0[A]\big(\overleftarrow{\pa}_{\!A}\big)^2h^2(A), ... ,\\
\label{D5}
&&\delta_{\xi}^{(1)}A=-h(A)\overleftarrow{\pa}_{\!A}R^{(0)}(A)\xi +
R^{(0)}(A)\overleftarrow{\pa}_{\!A}h(A)\xi,\\
\nonumber
&&\delta_{\xi}^{(2)}A=\big(h(A)\overleftarrow{\pa}_{\!A}\big)^2R^{(0)}(A)\xi -
\big(h(A)\overleftarrow{\pa}_{\!A}\big)\big(R^{(0)}(A)\overleftarrow{\pa}_{\!A}h(A)\big)\xi
+\frac{1}{2}R^{(0)}(A)\big(\overleftarrow{\pa}_{\!A}h(A)\big)^2\xi,... .
\eeq
For deformation of free models one has the following specifications
\beq
\label{D6}
S_n[A]=0, \; n=3,4,... ,\quad R^{(0)}(A)\overleftarrow{\pa}_{\!A}h(A)\xi =0,
\eeq
so that $ \delta_{\xi}^{(n)}A=\big(h(A)\overleftarrow{\pa}_{\!A}\big)^n R^{(0)}(A)\xi,
\; n=2,3,...$ . For anticanonical transformations used in the present paper there exists
additional restriction $h(A)\overleftarrow{\pa}_{\!A}=0$ leading to the absence of deformation
for the original gauge symmetry.
From above presentation, it follows evident advantages of the new method.
Note that in recent studies of the high spin theory expansions of
the deformed action and
symmetry are considered not in the deformation parameter but in fields
\cite{Zinov,JT,FKM,KMP}). Relation
between such Noether procedure and new method corresponds to expansion of generating function
$h(A)$ in the above equations (\ref{D5}), (\ref{D6}),
\beq
\label{D7}
h(A)=h_2(A)+h_3(A)+\cdots ,
\eeq
where $h_k(A)$, $(k=2,3,..., )$  is a function of $k$-th order in fields providing then by
corresponding reorganization of Taylor series.
In particular, the triviality of deformed gauge symmetry claimed in
\cite{JT,FKM}) when cubic vertices are vanished  is formulated in the new method as
\beq
h(A)\overleftarrow{\pa}_{\!A}R^{(0)}(A)\xi=0, \qquad
R^{(0)}(A)\overleftarrow{\pa}_{\!A}h(A)\xi=0,
\eeq
without any reference to cubic vertices.

On the first sight, it seems that the obtained result (\ref{c16})
contradicts with the result
of paper \cite{Zinov}.\footnote{ Notice that in \cite{BIZ2} $N=2$
supersymmetric generalization of cubic vertices containing in bosonic sector interactions
of spin 0 field with massless integer spin $s$ fields has been considered.} The
corresponding action $S_{1 Zin}=S_{1 Zin}[\varphi,\phi]$ in \cite{Zinov} has the form
\beq
\label{D8}
S_{1 Zin}=\frac{a_0}{6}\varepsilon^{ij}\int dx
\varphi_{\mu\nu\lambda}\Big[-2\pa^{\mu}\pa^{\nu}\phi_i\;
\pa^{\lambda}\phi_j+2\eta^{\nu\lambda}\Box\phi_i\;\pa^{\mu}\phi_j+
\eta^{\nu\lambda}\pa^{\mu}\pa_{\sigma}\phi_i\;\pa^{\sigma}\phi_j\Big].
\eeq
Indeed, this action is not invariant under the gauge transformations (\ref{c2}),
\beq
\label{D9}
\delta S_{1 Zin}=-\frac{a_0}{6}\varepsilon^{ij}\int dx
\xi_{\nu\lambda}\Big[-2\pa^{\sigma}\pa^{\nu}\phi_i\;\pa^{\lambda}\pa_{\sigma}\phi_j
-2 \pa^{\nu}\pa^{\lambda}\phi_i\;\Box\phi_j+
4\Box\phi_i\;\pa^{\nu}\pa^{\lambda}\phi_j\Big]\neq 0.
\eeq
But it needs to have in mind that the cubic vertices for the initial
free model of massless scalar fields and massless spin 3 fields
were studied using the standard Noether's procedure when scalar fields were
subjected to gauge transformations too. Then, the gauge invariance of the theory using
 action (\ref{D8}) is supported in the linear approximation only \cite{Zinov}.
In the approach applied in Section 4 it corresponds to
anticanonical transformations when the generating function responsible for change of
massless scalar
field $\phi$ ($m=0$) is not equal to zero, $h\neq 0$. In its turn,
it will lead to non-trivial
deformations  of scalar fields and gauge algebra.
So, there is no any contradictions between the result of  \cite{Zinov} and the
result obtained in the present paper because these tasks are different
in their formulation.

We can analyze the interactions of massless spin 3 field with the pair of
scalars $\phi^i,\;i=1,2$
within the new approach leading to  the cubic vertex
which will be  anti-symmetric on them.
The most general form of generating function containing three partial derivatives
 reads
\beq
\nonumber
&&h^{\mu\nu\lambda}=a_0\frac{1}{\Box}\varepsilon^{ij}
\big(c_1\pa^{\mu}\pa^{\nu}\pa^{\lambda}\phi_i \;\phi_j+
c_2\pa^{(\mu}\pa^{\nu}\phi_i\;\pa^{\lambda)}\phi_j+
c_3\eta^{(\mu\nu}\Box\pa^{\lambda)}\phi_i\;\phi_j
+\\
\label{D10}
&&\qquad\qquad\quad+c_4\Box\phi_i\;\eta^{(\mu\nu}\pa^{\lambda)}\phi_j+
c_5\eta^{(\mu\nu}\pa^{\lambda)}\pa_{\sigma}\phi_i\;\pa^{\sigma}\phi_j\big),
\eeq
where $a_0$ is the coupling constant with ${\rm dim}(a_0)=-d/2$ and $c_i,\;\; i=1.2,...,5$
are arbitrary dimensionless constants. Then, the local addition to the initial action
(\ref{c1}) has the form
\beq
\nonumber
&&S_{1\; loc}=2a_0\int dx \varphi_{\mu\nu\lambda}\varepsilon^{ij}
\Big[c_1\pa^{\mu}\pa^{\nu}\pa^{\lambda}\phi_i \;\phi_j+
c_2\pa^{(\mu}\pa^{\nu}\phi_i\;\pa^{\lambda)}\phi_j-\\
\nonumber
&&\qquad\qquad\qquad-
\big(c_1+c_3(d+1)+2c_4(d+1)\big)\eta^{(\mu\nu}\Box\pa^{\lambda)}\phi_i\;\phi_j
-\\
\nonumber
&&\qquad\qquad\qquad-\big(c_2+c_4(d+1)\big)\Box\phi_i\;\eta^{(\mu\nu}\pa^{\lambda)}\phi_j-\\
&&\qquad\qquad\qquad-
\big(2c_2+c_5(d+1)\big)\eta^{(\mu\nu}\pa^{\lambda)}\pa_{\sigma}\phi_i\;\pa^{\sigma}\phi_j\Big],
\label{D11}
\eeq
Due to the fact that the gauge algebra does not deform under anticanonical transformations with
the generating function (\ref{D10}), the functional (\ref{D11})
must be invariant under original
gauge transformations, $\delta S_{1\; loc}=0, \;
\delta\varphi_{\mu\nu\lambda}=\pa_{(\mu}\xi_{\nu\lambda)},\; \eta_{\mu\nu}\xi^{\mu\nu}=0,\;
\delta\phi_i=0,\; i=1,2$. Analysis of this requirement leads to the result
\beq
\nonumber
&&S_{1\; loc}=2a_0c_1\int dx \varphi_{\mu\nu\lambda}\varepsilon^{ij}
\Big[\pa^{\mu}\pa^{\nu}\pa^{\lambda}\phi_i \;\phi_j+
\pa^{(\mu}\pa^{\nu}\phi_i\;\pa^{\lambda)}\phi_j+\\
&&\qquad\qquad\qquad+
\frac{1}{2}\eta^{(\mu\nu}\Box\pa^{\lambda)}\phi_i\;\phi_j
 +\frac{1}{2}\Box\phi_i\;\eta^{(\mu\nu}\pa^{\lambda)}\phi_j
 -\eta^{(\mu\nu}\pa^{\lambda)}\pa_{\sigma}\phi_i\;\pa^{\sigma}\phi_j\Big],
\label{D12}
\eeq
which is different from (\ref{D8}).

We can also discuss the situation with quartic vertices, $\sim\varphi\phi\phi\phi$,
in the case under consideration. To construct such  interactions, in general,
one has to use the generating functions $h^{\mu\nu\lambda}$ in the sector of
fields $\varphi^{\mu\nu\lambda}$ and $h^i$ in the sector of fields $\phi^i$.
In turn, the structure of generating functions should be as follows
$h^{\mu\nu\lambda}\sim \varepsilon^{jk}\pa^{\mu}
\pa^{\nu}\pa^{\lambda} \phi_i\phi_j\phi_k$
and
$h^i\sim\varepsilon^{jk}\pa_{\mu}\pa_{\nu}\pa_{\lambda}\varphi^{\mu\nu\lambda}\phi_j\phi_k$
that is contrary to the tensor properties for these functions. Therefore,
quartic vertices are forbidden.

Quite recently, it was shown that non-trivial quartic vertices describing interactions among
massless spin 3 field, $\varphi^{\mu\nu\lambda}$, and massive vector, $A_{\mu}$,
and real scalar, $\phi$, fields of the form $\sim\varphi A \phi\phi$ can be constructed
\cite{L-22-2}. These vertices are invariant under original gauge transformations, similar
to the case considered in Section 4.

\section*{Acknowledgments}
\noindent
The author thanks I.L. Buchbinder for stimulating discussions of different aspects
concerning the problem of interactions for higher spin fields.
The work is supported by the Ministry of
Education of the Russian Federation, project FEWF-2020-0003.


\begin {thebibliography}{99}
\addtolength{\itemsep}{-8pt}

\bibitem{BV} I.A. Batalin, G.A. Vilkovisky, \textit{Gauge algebra and
quantization}, Phys. Lett. \textbf{B102} (1981) 27.

\bibitem{BV1} I.A. Batalin, G.A. Vilkovisky, \textit{Quantization of gauge
theories with linearly dependent generators}, Phys. Rev.
\textbf{D28} (1983)
2567.

\bibitem{BH}
G. Barnich, M. Henneaux, \textit{Consistent coupling between fields
with gauge freedom and deformation of master equation}, Phys. Lett.
{\bf B} 311 (1993) 123-129, {arXiv:hep-th/9304057}.

\bibitem{H}
M. Henneaux, \textit{Consistent interactions between gauge fields:
The cohomological approach}, Contemp. Math. {\bf 219} (1998) 93-110,
{arXiv:hep-th/9712226}.

\bibitem{BL-1}
I.L. Buchbinder, P.M. Lavrov,
\textit{On a gauge-invariant deformation of a classical gauge-invariant
theory}, JHEP 06 (2021) 854, {arXiv:2104.11930 [hep-th]}.


\bibitem{BL-2}
I.L. Buchbinder, P.M. Lavrov,
\textit{On classical and quantum deformations of gauge theories},
Eur. Phys. J. {\bf C} 81 (2021) 856, {arXiv:2108.09968 [hep-th]}.

\bibitem{L-2022}
P.M. Lavrov,
\textit{On gauge-invariant deformation of reducible
gauge theories}, Eur. Phys. J. {\bf C} 82 (2022) 429, {arXiv:2201.07505 [hep-th]}.


\bibitem{BFV1}
E.S. Fradkin, G.A. Vilkovisky, \textit{Quantization of relativistic
systems with constraints}, Phys. Lett. {\bf B} 55 (1975) 224.

\bibitem{BFV2}
I.A. Batalin, G.A. Vilkovisky, \textit{Relativistic S-matrix of
dynamical systems with boson and fermion constraints}, Phys. Lett.
{\bf B} 69 (1977) 309.

\bibitem{BFV3}
I.A. Batalin, E.S. Fradkin, \textit{Operator quantization of
relativistic dynamical system subject to first class constraints},
Phys. Lett. {\bf B} 128 (1983) 303.


\bibitem{BL-3}
I.L. Buchbinder, P.M. Lavrov,
\textit{On deformations of constrained Hamiltonian systems in BFV-formalism},
{arXiv:2203.05313 [hep-th]}.

\bibitem{BLT}
I.A. Batalin, P.M. Lavrov, I.V. Tyutin,
{\it Finite anticanonical transformations in field-antifield formalism},
Eur. Phys. J. {\bf C75} (2015) 270.

\bibitem{BLT-21}
I.A. Batalin, P.M. Lavrov, I.V. Tyutin,
{\it Anticanonical transformations and Grand Jacobian},
Russ. Phys. J. {\bf 64} (2021) 688.

\bibitem{VLT}
B.L. Voronov, P.M. Lavrov, I.V. Tyutin,
{\it Canonical transformations and gauge dependence
in general gauge theories},
Sov. J. Nucl. Phys. {\bf 36} (1982) 292.

\bibitem{GWen}
J. Gomis, S. Weinberg,
{\it Are nonrenormalizable gauge theories renormalizable?},
Nucl. Phys. B {\bf 469} (1996) 473.

\bibitem{Fronsdal-1}
C. Fronsdal, {\it Massless field with integer spin}, Phys. Rev. {\bf
D18} (1978) 3624.

\bibitem{BGK}
I.L. Buchbinder, A.V. Galajinsky, V.A. Krykhtin,
{\it Quartet unconstrained formulation for massless higher spin fields},
Nucl. Phys. {\bf B779} (2007) 155.

\bibitem{BG}
I.L. Buchbinder, A.V. Galajinsky,
{\it Quartet unconstrained formulation for massive higher spin fields},
JHEP 11 (2008) 081, {arXiv: 0810.2852 [hep-th]}.

\bibitem{BIZ}
I. Buchbinder, E. Ivanov, N. Zaigraev,
{\it Unconstrained off-shell superfield formulation of 4D,
N = 2 supersymmetric higher spins},
JHEP 12 (2021) 016, {arXiv:2109.07639 [hep-th]}.


\bibitem{DeWitt}
B.S. DeWitt, \textit{Dynamical theory of groups and fields},
(Gordon and Breach, 1965).

\bibitem{Zinov}
Yu.M. Zinoviev, \textit{Spin 3 cubic vertices in a frame-like formalism},
JHEP 08 (2010) 084, {arXiv:1007.0158 [hep-th]}.

\bibitem{JT}
E. Joung, M. Taronna,
\textit{A note on higher-order vertices of higher-spin fields
in flat and (A)dS space},
 JHEP 09 (2020) 171, arXiv:1912.12357 [hep-th].

 \bibitem{FKM}
S. Fredenhagen, O. Kr\"{u}ger, K. Mkrtchyan,
\textit{Restrictions for nn-Point Vertices in Higher-Spin Theories},
JHEP 06 (2020) 118, arXiv:1912.13476 [hep-th].

\bibitem{KMP}
M. Karapetyan, R. Manvelyan, G. Poghosyan,
\textit{On special quartic interaction of higher spin gauge fields
with scalars and gauge symmetry commutator in the linear approximation},
Nucl. Phys.B 971 (2021) 115512, arXiv:2104.09139 [hep-th].

\bibitem{BIZ2}
I. Buchbinder, E. Ivanov, N. Zaigraev,
{\it Off-shell cubic hypermultiplet couplings to N=2 higher spin gauge superfields},
JHEP 05 (2022) 104, {arXiv:2202.08196 [hep-th]}.

\bibitem{L-22-2}
P.M. Lavrov,
\textit{Gauge-invariant models of interacting fields with spins 3, 1 and 0}.
{arXiv:2209.03678 [hep-th]}.


\end{thebibliography}
\end{document}